\documentclass[conference]{IEEEtran}
\IEEEoverridecommandlockouts
\usepackage{cite}
\usepackage{amsmath,amssymb,amsfonts}
\usepackage{algorithm}
\usepackage{algorithmic}
\usepackage{graphicx}
\usepackage{subcaption}
\usepackage{textcomp}
\usepackage{xcolor}
\makeatletter

\newcommand{\Rmnum}[1]{\expandafter\@slowromancap\romannumeral #1@}
\makeatother
\def\BibTeX{{\rm B\kern-.05em{\sc i\kern-.025em b}\kern-.08em
    T\kern-.1667em\lower.7ex\hbox{E}\kern-.125emX}}
     
\title{UAV-Aided Progressive Interference Source Localization Based on Improved Trust Region Optimization\\
\thanks{This work was supported in part by the National Natural Science Foundation of China under Grants No. 6240012897 and 62271250, in part by the Jiangsu Provincial Natural Science Foundation of China under Grant No. SBK2024047654, and in part by the Key Technologies R\&D Program of Jiangsu (Prospective and Key Technologies for Industry) under Grants BE2022067, BE2022067-1, BE2022067-3.}}

\author{\IEEEauthorblockN{Guochen Gu$^{1}$, Zhipeng Lin$^{2}$, Qiuming Zhu$^{2}$, Junchang Chen$^{4}$,\\ Qihui Wu$^{2}$, Hongtao Duan$^{5}$, Yang Huang$^{2}$, Weizhi Zhong$^{3}$}
\IEEEauthorblockA{
\textit{$^{1}$College of Computer Science and Technology, Nanjing University of Aeronautics and Astronautics}, Nanjing, China\\
 $^{2}$\textit{College of Electronic and Information Engineering, Nanjing University of Aeronautics and Astronautics}, Nanjing, China\\
 $^{3}$\textit{College of Astronautics, Nanjing University of Aeronautics and Astronautics}, Nanjing, China\\
 $^{4}$\textit{National Radio Monitoring Center Yunnan Monitoring Station}, Chengjiang, China\\
 $^{5}$\textit{State Radio Monitoring Center}, Beijing, China\\
Email: linlzp@nuaa.edu.cn}
}
\begin{document}
\maketitle
\begin{abstract}
Trust region optimization-based received signal strength indicator (RSSI) interference source localization methods have been widely used in low-altitude research. However, these methods often converge to local optima in complex environments, degrading the positioning performance. This paper presents a novel unmanned aerial vehicle (UAV)-aided progressive interference source localization method based on improved trust region optimization. By combining the Levenberg-Marquardt (LM) algorithm with particle swarm optimization (PSO), our proposed method can effectively enhance the success rate of localization. We also propose a confidence quantification approach based on the UAV-to-ground channel model. This approach considers the surrounding environmental information of the sampling points and dynamically adjusts the weight of the sampling data during the data fusion. As a result, the overall positioning accuracy can be significantly improved. Experimental results demonstrate the proposed method can achieve high-precision interference source localization in noisy and interference-prone environments.
\end{abstract}

\begin{IEEEkeywords}
Interference Source Localization, UAV, RSSI, Data fusion, Confidence Quantification
\end{IEEEkeywords}

\section{Introduction}
In low-altitude research, accurate localization of interference sources is essential to maintain the reliability of communication, navigation, and sensing systems in congested electromagnetic environments \cite{b1}. Unmanned aerial vehicles (UAVs) offer an efficient solution by providing flexible coverage and overcoming signal obstructions from buildings \cite{b13}. Among numerous localization techniques, received signal strength indicator (RSSI)-based methods have gained significant attention due to their simplicity and cost-effectiveness \cite{b2}. However, most existing RSSI-based localization methods often cannot achieve high accuracy in complex environments, as signal attenuation, reflection, and noises can significantly degrade positioning performance \cite{b3},\cite{b11}.

Trust region optimization-based methods have been proposed to address the limitations of RSSI-based localization. The authors in \cite{b4} proposed an extended multi-lateration method integrated with boundary consideration, zone selection, and estimated position compensation based on virtual positions to increase localization accuracy. A novel range-free weighted centroid approach for sensor node localization based on RSSI was proposed in \cite{b5}, which relies on topology and network information to estimate location in low communication traffic. However, a common drawback of such methods is their susceptibility to local optima, which can hinder their ability to achieve global accuracy.

Data fusion has been explored to enhance localization accuracy. In \cite{b6}, a filtered RSSI and beacon weight approach using Kalman filter achieved an accuracy of centimeters localization using cost-effective and easy-to-deploy beacons. In \cite{b7}, the authors considerably improved the position estimation error by adjusting the information-sharing coefficients online using a federated Kalman filter (FKF). However, most of these methods fail to account for the impact of sampling positions and the surrounding environment on localization accuracy. Thus, their positioning accuracy is still not very high.

In this paper, we propose a novel UAV-aided interference source localization method based on improved trust region optimization. Progressive localization is employed to adaptively refine the target position estimate by iteratively optimizing the sampling positions, enhancing the positioning accuracy and robustness. 

The contributions of this paper are as follows:
\begin{itemize}
\item An RSSI-driven interference source localization method based on improved trust region optimization is proposed. By combining the particle swarm optimization (PSO) with the Levenberg-Marquardt (LM) algorithm, the method enhances the localization success rate by avoiding local optima.
\item An improved weighted Kalman filter (KF) data fusion approach based on confidence quantification is designed. A confidence scaling function evaluates RSSI reliability and dynamically adjusts the weight factors in data fusion. This approach enhances localization accuracy and robustness, performing well under sophisticated environments.
\item Comprehensive simulation experiments are conducted by using a dataset generated from a real-world campus scenario. Simulation results demonstrate that the proposed progressive localization method can improve the positioning accuracy by about 60$\%$ over the prior art.
\end{itemize}

The rest of the paper is organized as follows. Section \Rmnum{2} introduces the system model and problem formulation. Section \Rmnum{3} presents the proposed interference source localization method. Section \Rmnum{4} provides the simulation results and analysis. Finally, conclusions are drawn in Section \Rmnum{5}.

\section{System Model and Problem Formulation}

\begin{figure}[t]
  \centering
\includegraphics[width=1\columnwidth]{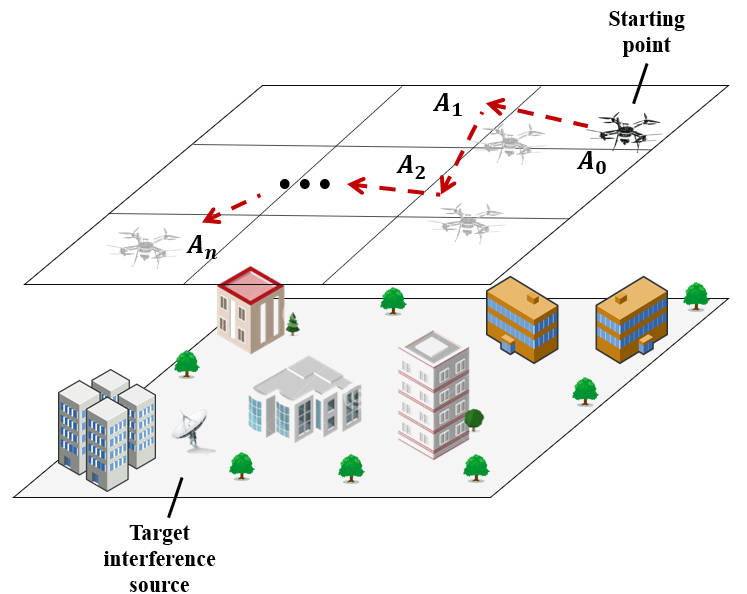}
  \caption{Interference source positioning scenario.}
  \label{fig:method}
\end{figure}

The region of interest (ROI) contains a target interference source with an unknown emission power coefficient $\eta$ and an unknown location $\theta = (\theta _{x},\theta _{y},\theta _{z})$. As shown in Fig. \ref{fig:method}, a UAV equipped with a spectrum sensing device is adapted to collect the RSSI data along the flight trajectory in the ROI. Assume that the UAV travels from point $A_0$ to point $A_1$, and sequentially collects RSSI measurements $R_{0, i}$ at specific spatial coordinates $(x_i,y_i,z_i), i = 1,2, \cdots, n$. These measurements form two corresponding datasets: the RSSI dataset $\mathbf{R_0}$ and the coordinate dataset $\mathbf{C_0}$, which represents the three-dimensional spatial positions of the UAV at the time of each measurement.

Mathematically, $\mathbf{R_0}$ and $\mathbf{C_0}$ can be expressed as follows:
\begin{align}
    \mathbf{R_0} = \{ R_{0,1}, R_{0,2}, \dots, R_{0,n} \},
\end{align}
\small
\begin{align}
    &\mathbf{C} = \nonumber \\
    &\{ (x_{0,1}, y_{0,1}, z_{0,1}), (x_{0,2}, y_{0,2}, z_{0,2}), \dots, (x_{0,n}, y_{0,n}, z_{0,n}) \},
\end{align}
\normalsize
where $n$ represents the total number of sampling points along the UAV's trajectory.

In this study, an enhanced channel propagation loss model that extends the conventional close-in (CI) path loss model by incorporating the impact of UAV altitude on the path loss exponent (PLE) is adopted  \cite{b8}. This model retains the physical characteristics of the CI model while emphasizing the critical role of PLE in capturing the attenuation characteristics of the propagation environment. The mathematical formulation of the proposed model is expressed as
\begin{equation}
L = 32.4+20\log_{10}\!\left(f_{\mathrm{c}}\right)
    +10\!\left(A\cdot h_{\mathrm{UAV}}^B\right)\!\log_{10}\!\left(d_i\right)
    +\chi_{\sigma}.
\label{eq:PLE}
\end{equation}
  
In which
\begin{align}
d_i=  \sqrt{\left(\theta_{x}-x_i\right)^2+\left(\theta_{y}-y_i\right)^2+\left(\theta_{z}-z_i\right)^2},
\end{align}
indicates the distance between the candidate source position and the sampling position; $L$ represents the PLE by the signal from the interference source to point $A_i$; the parameters $A$ and $B$ are environment-related constants; $\chi_{\sigma}$ represents a zero-mean Gaussian random variable accounting for shadow fading; $f_{\mathrm{c}}$ represents the carrier frequency (in GHz). This UAV-to-ground model is empirical, and the data used for fitting the model can be obtained either from field measurements or from ray tracing simulations.

Based on the equation \eqref{eq:PLE}, the relationship between the RSSI data collected at point $A_i$ and the distance $d_i$ can be derived as
\begin{align}
    R_{i}=P_{\eta}-10\left(A\cdot {h_{\mathrm{UAV}}}^{B}\right)\cdot\log_{10}\left(d_i\right)-\chi_{\sigma,i}
    \label{eq:R_i},
\end{align}
where $P_{\eta}$ is an unknown constant related to the emission power coefficient $\eta$ of the interference source.

Equation \eqref{eq:R_i} can be rewritten as 
\begin{align}
\mathbf{R}=P_{\eta}-\kappa\cdot\log_{10}\left(\mathbf{D}\right)-\chi_{\sigma},
\end{align}
where 
\begin{align}
    \kappa = 10\left(A\cdot {h_{\mathrm{UAV}}}^{B}\right),
\end{align}
\begin{align}
\mathbf{D}=\left[d_1,\cdots,d_n\right]\in\mathbb{R}^{n}.
\end{align}

In this case, the estimation of the interference source position is modeled as a linear regression problem, where $\chi_{\sigma}$ is modeled as a Gaussian distribution, $\mathbf{R}$ is known, $\mathbf{D\left( \cdot\right)}$ is a function of $\theta$, and $\theta$ is the parameter to be estimated. The positioning problem formulation can be expressed as
\begin{align}
\begin{array}{cl}{\text { Find }} & {\theta} \\ {\text { subject to }} & {\mathbf{R}=P_{\eta}-\kappa\cdot\log_{10}\left(\mathbf{D}\right)-\chi_{\sigma}}.
\label{eq:eta}
\end{array}
\end{align}

To address the aforementioned problem, this paper proposes a progressive interference source localization method, which operates iteratively through the following three main steps:
\begin{itemize}
\item \textbf{Trust region optimization-based localization:} In the $i$-th iteration, the RSSI data collected during the UAV's flight are used to derive a distance ratio array. A trust region optimization algorithm processes this array to obtain a preliminary target position estimate $\tilde{\mathbf{x}}_i$.
\item \textbf{Confidence-based data fusion:} The confidence $l_i$ of $\tilde{\mathbf{x}}_i$ is evaluated, and the target position is refined through a confidence-based data fusion approach, resulting in an updated position estimate $\mathbf{x}_i$.
\item \textbf{RSSI data sampling:} The UAV travels a distance $L$ from $A_i$ toward $\mathbf{x}_i$ to reach $A_{i+1}$, collecting RSSI data along the way.
\end{itemize}

These steps are repeated iteratively, starting with the UAV departing from a designated point $A_0$, flying a distance L in a random direction to $A_1$ for initial data collection. The process continues until the termination conditions are met, yielding the final precise target position.

\section{Proposed Positioning Method} 
\subsection{trust region optimization-based localization}

When the UAV travels from point $ A_i$ to point $A_{i+1}$, it evenly collects $n$ RSSI data along the way and generates an array $\mathbf{R_i}$.

It can be deduced from equation \eqref{eq:R_i} that
\begin{align}
    d_i=-10^{\frac{R_{i}-P_{\eta}}{10\left(A\cdot {h_{\mathrm{UAV}}}^{B}\right)}},
\end{align}
\begin{align}
    \frac{d_i}{d_j}=-10^{\frac{R_i-R_j}{10\left(A\cdot {h_{\mathrm{UAV}}}^{B}\right)}}.
    \label{eq:didj}
\end{align}

Using this formula, the difference in RSSI values between two sampling points can be transformed into the corresponding ratios of their distances to the target interference source. Based on this transformation, the array $\mathbf{R_i}$ is converted into a distance ratio array, denoted as 
\begin{align}
    d_{\text{ratio},i} = \{\frac{d_{i,1}}{d_{i,2}}, \frac{d_{i,1}}{d_{i,3}}, \dots, \frac{d_{i,1}}{d_{i,n}}\}.
\end{align}

Consequently, the problem \eqref{eq:eta} can be reformulated as a least squares optimization problem
\begin{align}
\begin{array}{cl}{\text { Find }} & {\tilde{\mathbf{x}}} \\ {\text { subject to }} & {\min\sum_{j=2}^{n} (\frac{\tilde{d}_j(\tilde{\mathbf{x}})}{\tilde{d}_1(\tilde{\mathbf{x}})}-d_{\text{ratio},i}(j-1))^2 }.
\end{array}
\end{align}
where $\tilde{d}_j(\tilde{\mathbf{x}})$ indicates the distance between the estimated source position $\tilde{\mathbf{x}}$ and the $j$-th sampling position $A_{i,j}$ in the trajectory. 

The PSO-LM algorithm is employed to address this least-squares optimization problem. The PSO algorithm, leveraging its population-based search mechanism, can identify an initial solution close to the global optimum within a relatively large search space. This reduces the number of iterations required by the LM algorithm and effectively mitigates the risk of the LM algorithm converging to a local optimum. The specific steps of the algorithm are outlined as follows:

Randomly initialize a particle swarm $\mathcal{P} = \{ \mathbf{x}_1, \mathbf{x}_2, \dots, \mathbf{x}_N \}$, within the search space, where $N$ denotes the number of particles, and each particle $\mathbf{x}_t \in \mathbb{R}^3$ represents a potential target position. For each particle $\mathbf{x}_t$, randomly initialize its velocity $\mathbf{v}_t \in \mathbb{R}^3$ and evaluate its fitness value
\begin{align}
    f(\mathbf{x}_t) = \sum_{j=2}^{n} (\frac{\tilde{d}_j(\mathbf{x}_t)}{\tilde{d}_1(\mathbf{x}_t)}-d_{\text{ratio},i}(j-1))^2.
\end{align}

Set the personal best position of each particle as $\mathbf{p}_t^{\text{best}} = \mathbf{x}_t$, and determine the global best position $\mathbf{g}^{\text{best}} = \arg\min_{\mathbf{x}_t} f(\mathbf{x}_t)$.

For each particle $\mathbf{x}_t$, update its velocity and position using the following equations:
\begin{align}
    &\mathbf{v}_t^{(k+1)} = \nonumber \\
    &\omega\mathbf{v}_t^{(k)} + c_1 r_1 (\mathbf{p}_t^{\text{best}} - \mathbf{x}_t^{(k)}) + c_2 r_2 (\mathbf{g}^{\text{best}} - \mathbf{x}_t^{(k)}),
\end{align}
\begin{align}
    \mathbf{x}_t^{(k+1)} = \mathbf{x}_t^{(k)} + \mathbf{v}_t^{(k+1)},
\end{align}
where $\omega$ denotes the inertia weight; $c_1$ and $c_2$ are acceleration coefficients; $r_1, r_2 \sim U(0, 1)$ are uniformly distributed random numbers.

Evaluate the updated fitness value and update the personal and global best positions. 
\begin{align}
    \text{If } f(\mathbf{x}_t^{(k+1)}) < f(\mathbf{p}_t^{\text{best}}), \quad \mathbf{p}_t^{\text{best}} = \mathbf{x}_t^{(k+1)},
\end{align}
\begin{align}
    \text{If } f(\mathbf{p}_t^{\text{best}}) < f(\mathbf{g}^{\text{best}}), \quad \mathbf{g}^{\text{best}} = \mathbf{p}_t^{\text{best}}.
\end{align}

Repeat the above process until achieving convergence in the global best solution.

The global best solution $\mathbf{g}^{\text{best}}$ obtained from the PSO phase is used as the initial solution for the LM algorithm, denoted as $\mathbf{x}^{(0)} = \mathbf{g}^{\text{best}}
$.

Set the initial solution $\mathbf{x}^{(0)}$, the damping factor $\lambda > 0$, and the objective function $f(\mathbf{x}) = \frac{1}{2}\sum_{j=2}^{n}r_j(x)^2$, where $\mathbf{r}_j(x) = \frac{\tilde{d}_j(\mathbf{x})}{\tilde{d}_1(\mathbf{x})}-d_\text{ratio}(j-1)$ is the residual vector.

At each iteration, compute the Jacobian matrix $\mathbf{J} = \frac{\partial \mathbf{r}}{\partial \mathbf{x}}$, and the residual vector $\mathbf{r}(\mathbf{x})$.

Update the solution vector $\mathbf{x}$ as follows
\begin{align}
    \mathbf{x}^{(k+1)} = \mathbf{x}^{(k)} - (\mathbf{J}^T \mathbf{J} + \lambda \mathbf{I})^{-1} \mathbf{J}^T \mathbf{r}(\mathbf{x}^{(k)}).
\end{align}

Adjust the damping factor $\lambda$ based on the change in the objective function. If $f(\mathbf{x}^{(k+1)}) < f(\mathbf{x}^{(k)})$, decrease $\lambda$. Otherwise, increase $\lambda$.

Stop the iterations when $\| \Delta \mathbf{x} \| < \epsilon$.

The final solution $\tilde{\mathbf{x}}_i$ obtained after the PSO-LM optimization phase is considered the preliminary localization result of the target interference source in the $i$-th iteration.

\subsection{confidence-based data fusion}

In the trust region optimization-based localization phase, the position of the target interference source is estimated based on the RSSI data. However, due to the impact of shadow fading and measurement noise, the obtained RSSI differential data exhibit a certain degree of uncertainty. So, it is necessary to quantify the confidence level of the RSSI differentials and employ a data fusion approach to mitigate the effects of measurement uncertainty, thereby enhancing localization accuracy.

In the path loss model, shadow fading is modeled as Gaussian noise with zero mean. The RSSI measurement at point $A_i$ can be expressed as
\begin{align}
    R_i = R_{\text{true}}(d_i) + \varepsilon_i, \quad \varepsilon_i \sim \mathcal{N}(0, \sigma^2),
\end{align}
where $\varepsilon_i$ is the shadow fading noise with variance $\sigma^2$. 

For two sampling points at distances $d_i$ and $d_j$, the measured RSSI difference is
\begin{align}
    \Delta R_{\text{measured}} = R_i - R_j = \Delta R_{\text{true}} + \varepsilon_\Delta,
\end{align}
where $\varepsilon_\Delta = \varepsilon_i - \varepsilon_j$. 

Since $\varepsilon_i$ and $\varepsilon_j$ are independent Gaussian random variables, their difference $\varepsilon_\Delta$ is also Gaussian, with
\begin{align}
    \varepsilon_\Delta \sim \mathcal{N}(0, 2\sigma^2).
\end{align}

Thus, the measured RSSI difference follows the distribution
\begin{align}
    \Delta R_{\text{measured}} \sim \mathcal{N}(\Delta R_{\text{true}}, 2\sigma^2).
\end{align}

Based on formula \eqref{eq:didj}, the true RSSI difference between two points at distances $d_i$ and $d_j$ is
\begin{align}
    \Delta R_{\text{true}} = -10\left(A\cdot {h_{\mathrm{UAV}}}^{B}\right) \log_{10}\left(\frac{d_i}{d_j}\right).
\end{align}

The probability that the true RSSI difference lies within a range centered on the measured difference is computed. The confidence of the measured RSSI difference is given by
\begin{align}
    \text{Confidence} = \Phi\left( \frac{\Delta R_{\text{measured}} - \Delta R_{\text{true}}}{\sqrt{2}\sigma} \right), 
\end{align}
where $\Phi(\cdot)$ is the cumulative distribution function (CDF) of the standard normal distribution. The standard normal CDF is defined as
\begin{align}
    \Phi(x) = \frac{1}{\sqrt{2\pi}} \int_{-\infty}^{x} e^{-\frac{t^2}{2}} \, dt.
\end{align}

The confidence of the target position estimate obtained from the RSSI data collected from $A_i$ to $A_{i+1}$ is further given by
\begin{align}
    l_{i} =\frac{1}{n-1} \sum_{t=1}^{n-1}{\Phi(\frac{\Delta R_{\text{measured},i,t} - \Delta R_{\text{true},i,t}}{\sqrt{2}\sigma})}.
\end{align}

The Kalman filter algorithm, incorporating the confidence $l_{i}$ of the position estimation, is employed in each iteration of the proposed progressive localization algorithm. This algorithm fuses the target position estimation $\tilde{\mathbf{x}}_i$ obtained during the $i$-th flight of the UAV with the historical target position estimations $\mathbf{x}_{i-1}$.

Specifically, in the framework of the conventional KF algorithm, during the $i$-th fusion, $\mathbf{x}_{i-1}$ is regarded as a priori state estimate; $\tilde{\mathbf{x}}_i$ is regarded as a measurement state estimate; the measurement noise covariance $p_i$ is corrected to $p_i = \frac{p_0}{l_i} $, where $p_0$ is the reference noise covariance; the state prediction matrix $F_i$ and the measurement matrix $H_i$ are both unit matrices. The posterior state estimate $\mathbf{x}_{i}$ obtained by the $i$-th fusion is the target position obtained by the $i$-th iteration, which is also regarded as the priori state estimate in the $i+1$-th fusion. 

\section{Simulations and Analysis}
The performance of the proposed localization method is verified through extensive MATLAB simulations. In the simulations, the commercial ray tracing software, Wireless Insite is used to generate the RSSI data of the ROI. The simulations were based on a 3D model of a real-world campus scenario. The size of the ROI is $1250$m $ \times$ $1250$m. The frequency measured is $f_c=2450$ Hz. The UAV’s flight distance $L$ for each iteration is $100$m, and RSSI measurement is performed every $10$m. According to empirical values, parameters in the path loss model are  $A=2.772$, $B=-0.04724$, $\eta = 3.53$. The accuracy of the positioning result is defined as the root mean squared error (RMSE).

We compare the proposed progressive localization method with three other optimization-based localization methods: the PSO-LM method, which performs a single optimization localization step without iteration; the trilateration method, which leverages RSSI differences to establish geometric relationships among sets of three sampling points for localization \cite{b9}; and the iterative centroid localization method, which estimates the target position iteratively by calculating the weighted centroid of the sampling points \cite{b10}.

\begin{figure}[t]
  \centering
  \includegraphics[width=1\columnwidth]{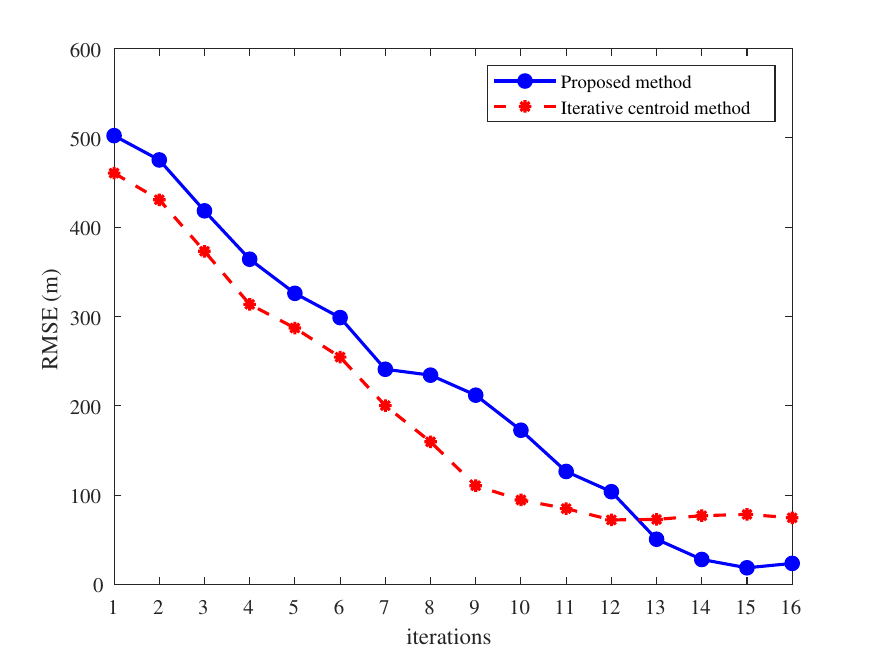}
  \caption{Positioning RMSE vs. number of iterations.}
  \label{fig:iterations}
\end{figure}

Fig. \ref{fig:iterations} illustrates the relationship between the positioning accuracy and the number of iterations at a signal-to-noise ratio (SNR) of 20 dB. It is evident that as the number of iterations increases, both two methods show a reduction in positioning error. However, after convergence, the proposed method achieves higher localization accuracy. This improvement can be attributed to the ability of the proposed method to mitigate the risk of converging to local optima and dynamically adjust the weight of RSSI data based on the surrounding environment of the sampling points during the iterative process. 

\begin{figure}[t]
  \centering
  \includegraphics[width=1\columnwidth]{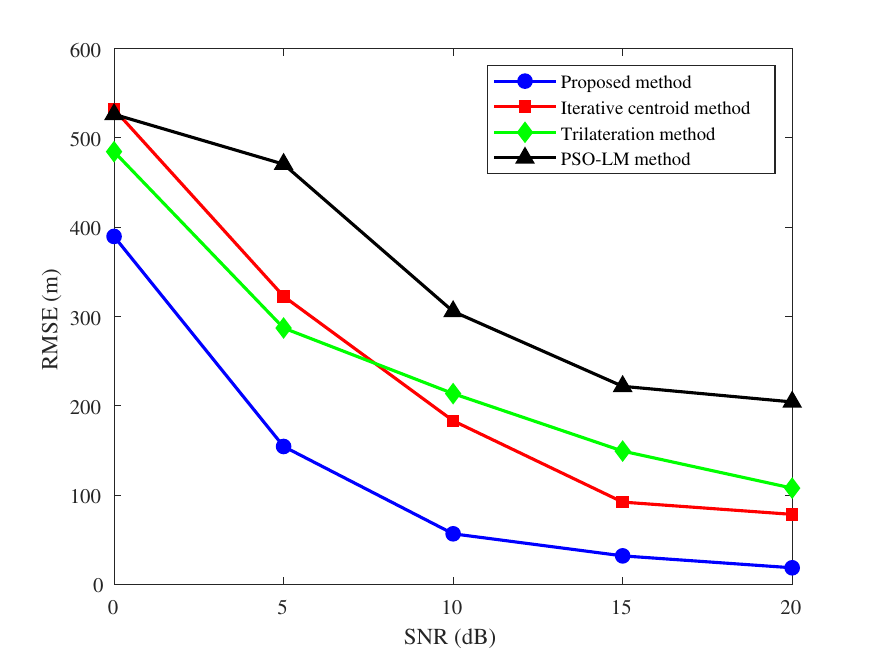}
  \caption{Positioning RMSE vs. SNR.}
  \label{fig:SNR}
\end{figure}

Fig. \ref{fig:SNR} depicts the relationship between positioning accuracy and SNR for each method. As the SNR increases from 0 dB to 20 dB, the RMSE of all target positioning results decreases, and the proposed method consistently outperforms the other three baseline methods.
This is because the proposed method is based on a more realistic UAV-to-ground channel model, which improves the precision of converting RSSI data into distance values. Additionally, the improved trust-region optimization method further boosts the positioning accuracy.

\begin{figure}[t]
  \centering
  \includegraphics[width=0.9\columnwidth]{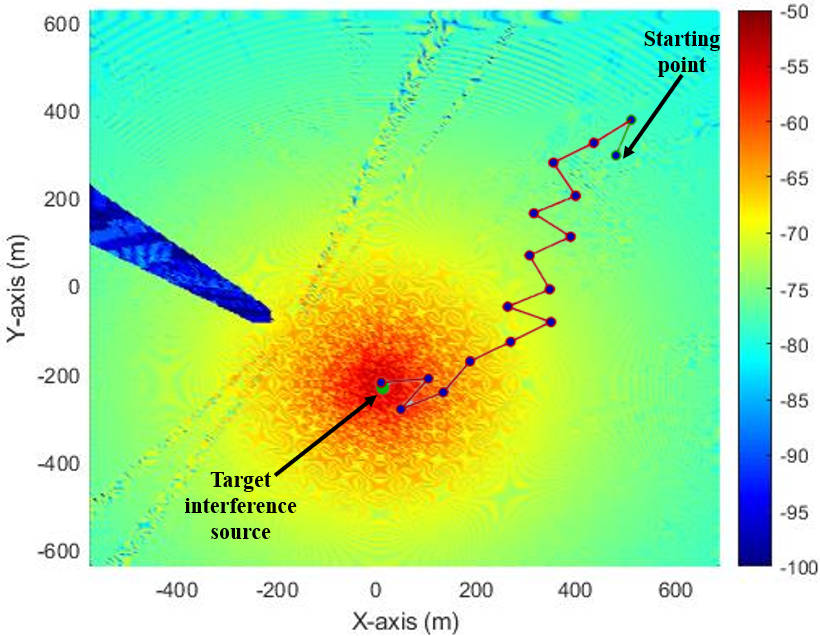}
  \caption{UAV trajectory during localization.}
  \label{fig:test}
\end{figure}

Fig. \ref{fig:test} shows the flight trajectory of the UAV when the proposed localization method is performed. It can be seen that through continuous iterations, the drone progressively approaches the target interference source, ultimately achieving an accurate and reliable positioning result.

\section{conclusion}

In this paper, a novel progressive interference source localization method utilizing RSSI data has been proposed. The proposed method has integrated an iterative optimization framework based on the PSO-LM algorithm, effectively mitigating local optima issues and significantly enhancing the localization success rate. An improved weighted KF data fusion approach has been introduced, incorporating a UAV-to-ground channel model-based confidence quantification mechanism that dynamically adjusts the influence of sampled RSSI data. This confidence-driven fusion approach has improved the accuracy and adaptability of the system in the presence of multipath effects and environmental noises. Simulation results have shown that the proposed method improves positioning accuracy by 60$\%$ compared to prior methods, offering a promising solution for high-precision interference source localization.

\bibliographystyle{IEEEtran}
\bibliography{refs}

\begin{thebibliography}{10}
\providecommand{\url}[1]{#1}
\csname url@samestyle\endcsname
\providecommand{\newblock}{\relax}
\providecommand{\bibinfo}[2]{#2}
\providecommand{\BIBentrySTDinterwordspacing}{\spaceskip=0pt\relax}
\providecommand{\BIBentryALTinterwordstretchfactor}{4}
\providecommand{\BIBentryALTinterwordspacing}{\spaceskip=\fontdimen2\font plus
\BIBentryALTinterwordstretchfactor\fontdimen3\font minus \fontdimen4\font\relax}
\providecommand{\BIBforeignlanguage}[2]{{%
\expandafter\ifx\csname l@#1\endcsname\relax
\typeout{** WARNING: IEEEtran.bst: No hyphenation pattern has been}%
\typeout{** loaded for the language `#1'. Using the pattern for}%
\typeout{** the default language instead.}%
\else
\language=\csname l@#1\endcsname
\fi
#2}}
\providecommand{\BIBdecl}{\relax}
\BIBdecl

\bibitem{b1}
C.~Huang, S.~Fang, H.~Wu \emph{et~al.}, ``Low-altitude intelligent transportation: System architecture, infrastructure, and key technologies,'' \emph{Journal of Industrial Information Integration}, vol.~42, p. 100694, 2024.

\bibitem{b13}
A.~Couturier and M.~A. Akhloufi, ``A review on absolute visual localization for {UAV},'' \emph{Robotics and Autonomous Systems}, vol. 135, p. 103666, 2021.

\bibitem{b2}
H.~P. Mistry and N.~H. Mistry, ``{RSSI} based localization scheme in wireless sensor networks: A survey,'' in \emph{2015 Fifth International Conference on Advanced Computing \& Communication Technologies}.\hskip 1em plus 0.5em minus 0.4em\relax IEEE, 2015, pp. 647--652.

\bibitem{b3}
N.~Chuku and A.~Nasipuri, ``{RSSI}-based localization schemes for wireless sensor networks using outlier detection,'' \emph{Journal of Sensor and Actuator Networks}, vol.~10, no.~1, p.~10, 2021.

\bibitem{b11}
J.~Wang, Q.~Zhu, Z.~Lin \emph{et~al.}, ``Sparse bayesian learning-based hierarchical construction for {3D} radio environment maps incorporating channel shadowing,'' \emph{IEEE Transactions on Wireless Communications}, vol.~23, no.~10, pp. 14\,560--14\,574, 2024.

\bibitem{b4}
A.~Booranawong, K.~Sengchuai, D.~Buranapanichkit \emph{et~al.}, ``{RSSI}-based indoor localization using multi-lateration with zone selection and virtual position-based compensation methods,'' \emph{IEEE access}, vol.~9, pp. 46\,223--46\,239, 2021.

\bibitem{b5}
K.~Akhil, K.~Seethalakshmi, and S.~Sinha, ``{RSSI} based positioning system for {WSN} with improved accuracy,'' in \emph{2021 3rd International Conference on Signal Processing and Communication (ICPSC)}.\hskip 1em plus 0.5em minus 0.4em\relax IEEE, 2021, pp. 325--329.

\bibitem{b6}
L.~Alsmadi, X.~Kong, K.~Sandrasegaran \emph{et~al.}, ``An improved indoor positioning accuracy using filtered {RSSI} and beacon weight,'' \emph{IEEE Sensors Journal}, vol.~21, no.~16, pp. 18\,205--18\,213, 2021.

\bibitem{b7}
T.~Ayabakan and F.~Kerestecio{\u{g}}lu, ``{RSSI}-based indoor positioning via adaptive federated kalman filter,'' \emph{IEEE Sensors Journal}, vol.~22, no.~6, pp. 5302--5308, 2021.

\bibitem{b8}
H.~Ni, Q.~Zhu, B.~Hua \emph{et~al.}, ``Path loss and shadowing for {UAV}-to-ground {UWB} channels incorporating the effects of built-up areas and airframe,'' \emph{IEEE Transactions on Intelligent Transportation Systems}, vol.~25, no.~11, pp. 17\,066--17\,077, 2024.

\bibitem{b9}
B.~Chen, J.~Ma, L.~Zhang \emph{et~al.}, ``Research progress of wireless positioning methods based on {RSSI},'' \emph{Electronics}, vol.~13, no.~2, p. 360, 2024.

\bibitem{b10}
K.~Akhil, K.~Seethalakshmi, and S.~Sinha, ``{RSSI} based positioning system for {WSN} with improved accuracy,'' in \emph{2021 3rd International Conference on Signal Processing and Communication (ICPSC)}.\hskip 1em plus 0.5em minus 0.4em\relax IEEE, 2021, pp. 325--329.

\end{thebibliography}

\end{document}